\documentclass[conference]{IEEEtran}
\IEEEoverridecommandlockouts
\usepackage{cite}

\usepackage{amsmath,amssymb,amsfonts}
\usepackage{algorithmic}
\usepackage{graphicx}
\usepackage{textcomp}
\usepackage{xcolor}
\usepackage{multirow}
\usepackage[export]{adjustbox}[2011/08/13]\usepackage[export]{adjustbox}[2011/08/13]

\makeatletter
\def\endthebibliography{%
  \def\@noitemerr{\@latex@warning{Empty `thebibliography' environment}}%
  \endlist
}
\makeatother

\def\BibTeX{{\rm B\kern-.05em{\sc i\kern-.025em b}\kern-.08em
    T\kern-.1667em\lower.7ex\hbox{E}\kern-.125emX}}
    
\begin{document}

\title{Style Transfer and Self-Supervised Learning Powered Myocardium Infarction Super-Resolution Segmentation\\
}

\author{\IEEEauthorblockN{1\textsuperscript{st} Lichao Wang}
\IEEEauthorblockA{\textit{Department of Computing} \\
\textit{Imperial College London}\\
l.wang22@imperial.ac.uk}
\and
\IEEEauthorblockN{2\textsuperscript{nd} Jiahao Huang}
\IEEEauthorblockA{\textit{National Heart and Lung Institute} \\
\textit{Imperial College London}\\
 j.huang21@imperial.ac.uk}
\and
\IEEEauthorblockN{3\textsuperscript{rd} Xiaodan Xing}
\IEEEauthorblockA{\textit{National Heart and Lung Institute} \\
\textit{Imperial College London}\\
London, UK \\
x.xing@imperial.ac.uk}
\and
\IEEEauthorblockN{4\textsuperscript{rd} Yinzhe Wu}
\IEEEauthorblockA{\textit{National Heart and Lung Institute} \\
\textit{Imperial College London}\\
London, UK \\
yinzhe.wu18@imperial.ac.uk}
\and
\IEEEauthorblockN{5\textsuperscript{rd} Ramyah Rajakulasingam}
\IEEEauthorblockA{\textit{National Heart and Lung Institute} \\
\textit{Imperial College London}\\
London, UK \\
ramyah.rajakulasingam05@imperial.ac.uk}
\and
\IEEEauthorblockN{6\textsuperscript{rd} Andrew D. Scott}
\IEEEauthorblockA{\textit{National Heart and Lung Institute} \\
\textit{Imperial College London}\\
London, UK \\
a.scott07@imperial.ac.uk}
\and
\IEEEauthorblockN{7\textsuperscript{rd} Pedro F Ferreira}
\IEEEauthorblockA{\textit{National Heart and Lung Institute} \\
\textit{Imperial College London}\\
London, UK \\
p.f.ferreira05@imperial.ac.uk}
\and
\IEEEauthorblockN{8\textsuperscript{rd} Ranil De Silva}
\IEEEauthorblockA{\textit{National Heart and Lung Institute} \\
\textit{Imperial College London}\\
London, UK \\
r.desilva@imperial.ac.uk}
\and
\IEEEauthorblockN{9\textsuperscript{rd} Sonia Nielles-Vallespin}
\IEEEauthorblockA{\textit{National Heart and Lung Institute} \\
\textit{Imperial College London}\\
London, UK \\
s.nielles-vallespin@imperial.ac.uk}
\and
\IEEEauthorblockN{10\textsuperscript{rd} Guang Yang}
\IEEEauthorblockA{\textit{National Heart and Lung Institute} \\
\textit{Imperial College London}\\
London, UK \\
g.yang@imperial.ac.uk}

}
\maketitle

\begin{abstract}
This study proposes a pipeline that incorporates a novel style transfer model and a simultaneous super-resolution and segmentation model. The proposed pipeline aims to enhance diffusion tensor imaging (DTI) images by translating them into the late gadolinium enhancement (LGE) domain, which offers a larger amount of data with high-resolution and distinct highlighting of myocardium infarction (MI) areas. 
Subsequently, the segmentation task is performed on the LGE style image. An end-to-end super-resolution segmentation model is introduced to generate high-resolution mask from low-resolution LGE style DTI image. Further, to enhance the performance of the model, a multi-task self-supervised learning strategy is employed to pre-train the super-resolution segmentation model, allowing it to acquire more representative knowledge and improve its segmentation performance after fine-tuning. https: github.com/wlc2424762917/Med\_Img

\end{abstract}


\section{Introduction}
Diffusion tensor (DT) cardiovascular magnetic resonance (CMR) is a novel noninvasive tool that enables inference of sheetlet orientations, which are altered under pathological conditions \cite{teh2017validation}. Preliminary studies have demonstrated the potential of DT CMR to detect microstructural abnormalities in myocardial infarction (MI) \cite{khalique2020diffusion}, suggesting the feasibility of MI segmentation on diffusion tensor imaging (DTI) images. 

\begin{figure}[h]
\vspace{-1.8ex}
\centerline{\includegraphics[width=0.44\textwidth]{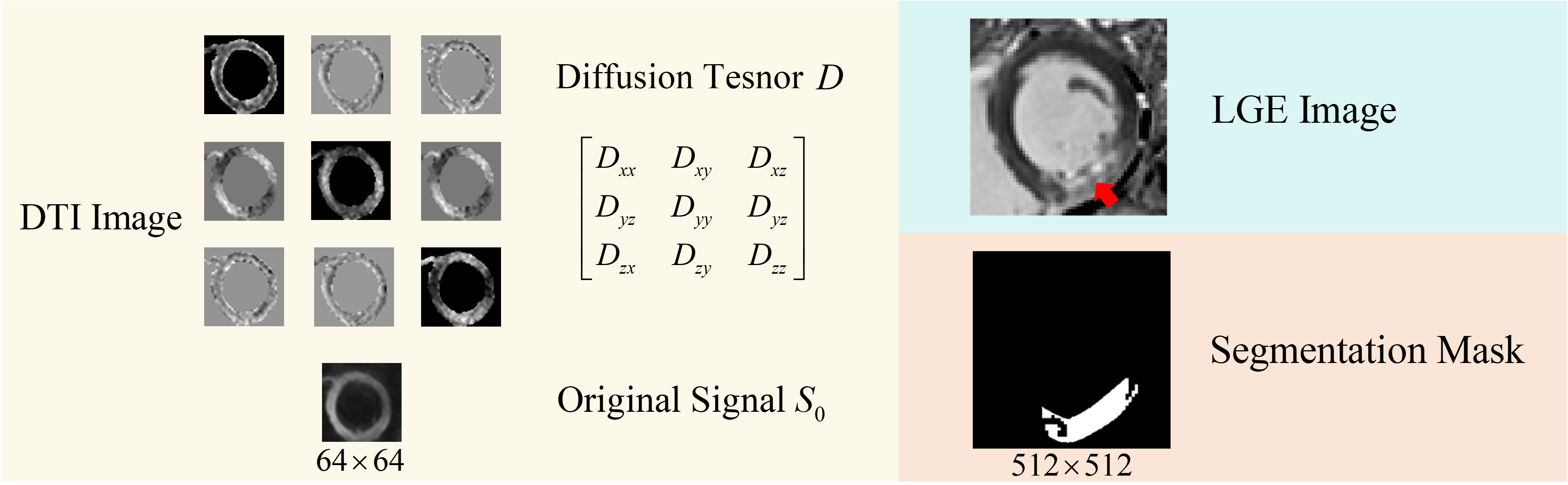}}
\vspace{-1.8ex}
\caption{Visualized data example of diffusion tensor imaging (DTI) and Late gadolinium enhancement (LGE) images. For DTI image, different components including DT $D$ and the corresponding original signal $S_0$ are presented. 
}

\vspace{-1ex}

\label{fig:data illustration}

\end{figure}
However, the domain of DTI has not been previously explored for MI segmentation. Current researches \cite{chen2022automatic, xu2022bmanet,yang2021hybrid,wang2020multi} were predominantly conducted using late gadolinium enhancement (LGE) images, due to certain limitations of DTI image compared to LGE image, including the lack of direct MI indication, a comparatively lower resolution, and a dearth of labeled data, as demonstrated in Fig. \ref{fig:data illustration}.

\begin{figure}[h]

\centerline{\includegraphics[width=0.4\textwidth]{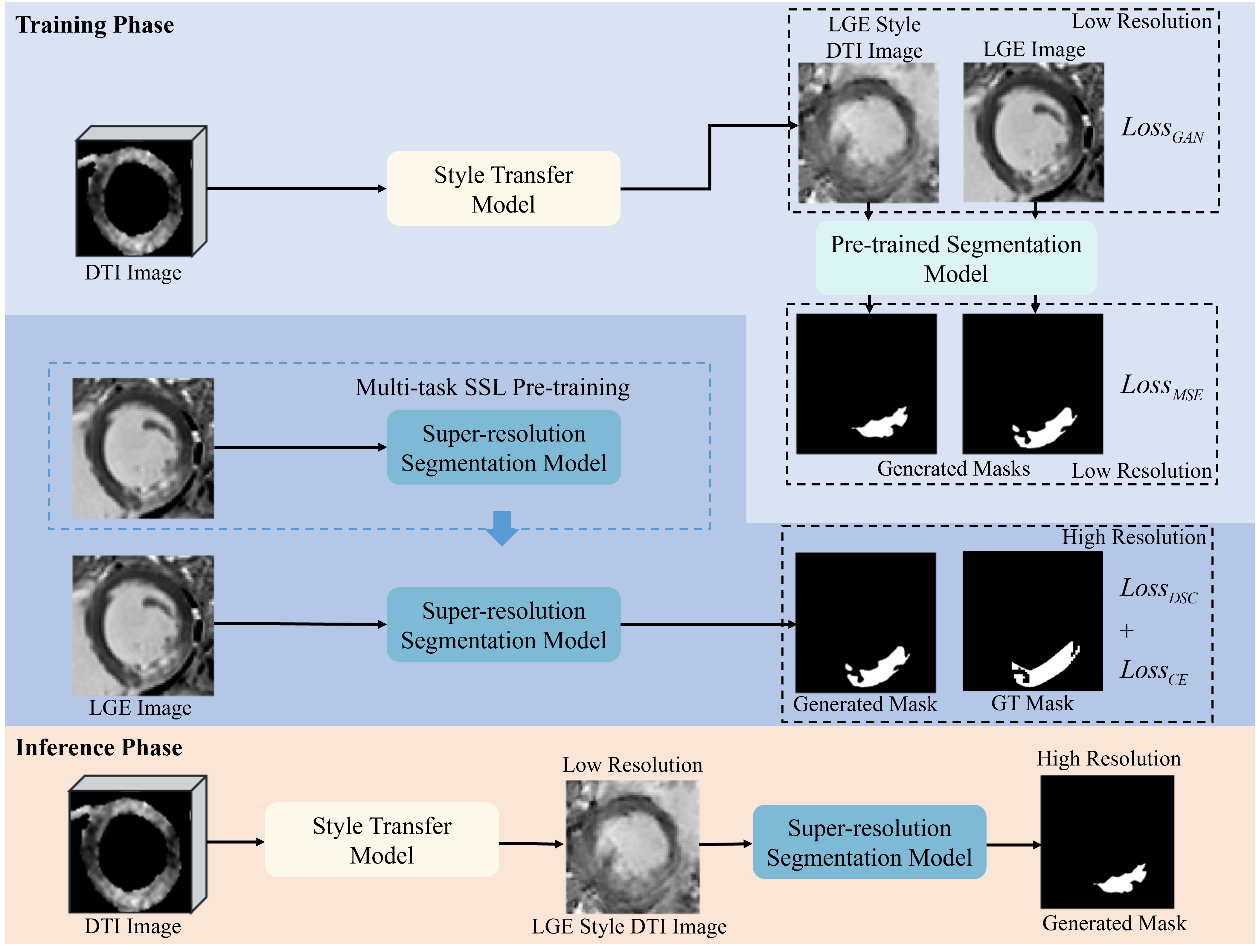}}
\vspace{-2ex}
\caption{The proposed pipeline. During the training phase, the style transfer model undergoes unsupervised training, while the super-resolution segmentation model is pre-trained using self-supervised pre-training and supervised fine-tuning strategy. In the inference phase, the Segmentor is excluded, and the pipeline is integrated. $Loss_{GAN}$, $Loss_{MSE}$, $Loss_{DSC}$, $Loss_{CE}$ stand for the CycleGAN loss \cite{zhu2017unpaired}, mean square error loss, dice loss, and cross-entropy loss respectively.}
\label{fig:pipeline}
\vspace{-3.3ex}

\end{figure}

To this end, we propose a novel pipeline to leverage the advantages of LGE data for MI segmentation on DTI image. The pipeline consists of a style transfer model and a super-resolution segmentation (SSeg) model, as depicted in Fig.\ref{fig:pipeline}. The style transfer model first converts the DTI image into LGE style, where MI areas are highlighted, and more labeled data is available. Then the SSeg model directly generates the high-resolution segmentation mask, facilitating improved segmentation of small foreground regions. 


Our style transfer model is based on the CycleGAN \cite{zhu2017unpaired}. To preserve the underlying segmentation mask throughout the style transfer, we develop the CycleGAN with a segmentation sub-network model, namely CycleGANSeg, aiming at keeping the integrity of the underlying MI segmentation mask.

Inspired by the Dual Super-Resolution Learning framework \cite{Wang2020DualSL}, we propose our SwinTransformer \cite{Liu2021SwinTH} based end-to-end SSeg model, namely SwinSSegNet. The SwinSSegNet aims to leverage more detailed information in the training process, and directly generates high-resolution segmentation masks in the inference process. 

Moreover, despite the relatively higher availability of LGE images, the quantity remains limited. To address this constraint, we construct a hybrid dataset by incorporating LGE images obtained from publicly available datasets, ACDC \cite{bernard2018deep} and LiVScar \cite{Mansi2015StatisticalAA}. Based on the hybrid dataset, we adopt a multitask self-supervised learning (SSL) pre-training strategy, including contrastive learning \cite{Chen2020ASF}, masked image modeling \cite{he2022masked}, and rotation prediction \cite{Gidaris2018UnsupervisedRL}. This approach enables our model to acquire a broad range of representation knowledge of LGE images, thereby enhancing its performance after fine-tuning. In summary, our contributions are as following:

1. We introduce a novel pipeline powered by style tansfer and SSL for MI super-resolution segmentation on DTI image. This pipeline effectively harnesses the abundant LGE data, and enhance the accuracy of MI segmentation on DTI image.

2. We present the CycleGANSeg, which incorporates a segmentation sub-network within the CycleGAN framework. This model is capable of converting DTI image to the LGE style, while preserving the integrity of the MI region. Notably, the CycleGANSeg can be trained in an unsupervised manner.

3. We propose the SwinSSegNet, an end-to-end super-resolution segmentation model that can directly generate high-resolution segmentation mask. This end-to-end paradigm has superior performance comparing to utilizing the 2-stage (first segmentation, then up-sample) paradigm.

4. We design a 2D multi-task self-supervised learning pre-training strategy on our curated hybrid LGE dataset to further enhance the performance of the SwinSSegNet model.

\section{Methods}
As shown in Fig \ref{fig:pipeline}, our proposed pipeline contains two models, a style transfer model, i.e., CycleGANSeg, and an SSeg model, i.e., SwinSSegNet. The CycleGANSeg initially translates the DTI image into the LGE style, with a resolution of $64\times64$. Subsequently, the SwinSSegNet generates the MI segmentation mask with an upsampled resolution of $512\times512$.

\subsection{CycleGAN with Segmentor}
\begin{figure}[h]
\vspace{-2ex}
\centerline{\includegraphics[width=0.46\textwidth]{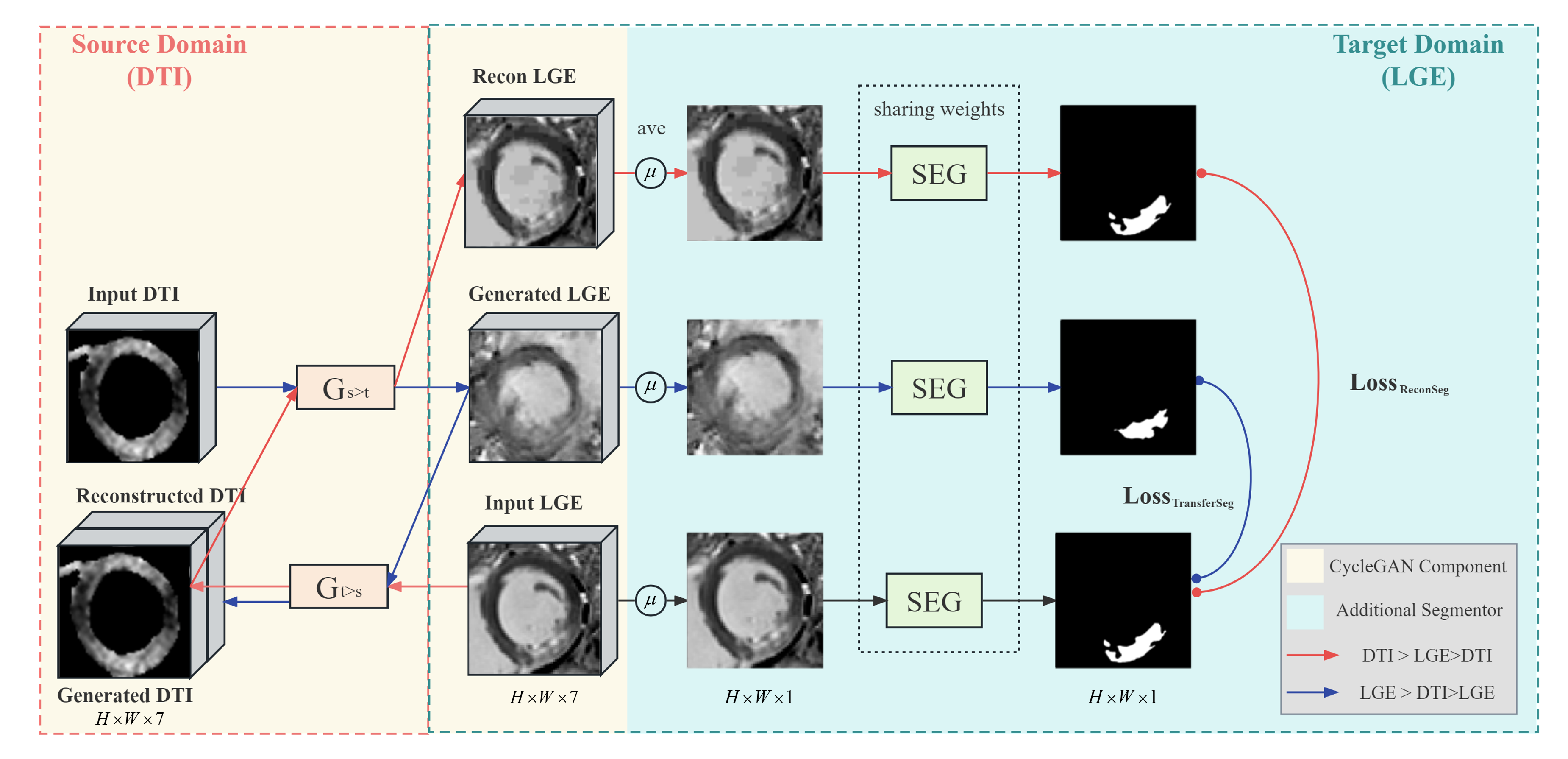}}
\vspace{-2ex}
\caption{The dataflow of CycleGANSeg. $G_{S>T}$, $G_{T>S}$, and SEG indcate the generator from DTI image to LGE image, the generator from LGE image to DTI image, and the Segmentor respectively. The black arrows, red arrows, and blue arrows depict the flow of data within the LGE domain, the bidirectional flow from LGE to DTI and back to LGE, and the bidirectional flow from DTI to LGE and back to DTI, respectively.}
\label{fig:CycleGANSeg}
\vspace{-1ex}
\end{figure}

As shown in Fig \ref{fig:CycleGANSeg}, the style transfer model is based on the CycleGAN. To keep the underlying segmentation mask unchanged in the style transfer process, a pre-trained segmentation network, namely Segmentor, is incorporated, along with two specific loss functions.

For the input and output data settings, we use the paired DTI images and LGE images as input, to ensure they have the same MI segmentation mask. Since our proposed CycleGANSeg comprises three sub-networks (a generator, a discriminator, and the Segmentor), the computational cost escalates significantly, as the resolution increases. Hence, to maintain efficiency during training, we standardize the resolution of all inputs and outputs to $64\times64$. 

The Segmentor outputs the segmentation mask of the original LGE image, the generated LGE image, and the reconstructed LGE image. To keep the integrity of the underlying segmentation mask, two mean square error (MSE) losses are adopted, as shown in \eqref{eq:loss_TransferSeg} and \eqref{eq:loss_ReconSeg}. This approach only requires paired images from different domains, therefore the CycleGANSeg offers the advantage of unsupervised training, obviating the need for extensive annotation. 

\vspace{-3ex}
\begin{equation}
    Loss_{TransferSeg} = \left ( Mask_{trans} - {Mask_{ori}} \right )  ^2,
    \label{eq:loss_TransferSeg}
\end{equation}
\vspace{-1.2ex}
\begin{equation}
     Loss_{ReconSeg} = \left ( Mask_{recon} - {Mask_{ori}} \right )  ^2,
    \label{eq:loss_ReconSeg}
\end{equation}
where $Mask_{trans}$, $Mask_{recon}$, and $Mask_{ori}$ stand for the mask of the generated LGE style DTI image, the reconstructed LGE image, and the original LGE image respectively.

\subsection{SwinTransformer Super-resolution Segmentation Model}
The SwinSSegNet is designed to directly generate the high-resolution segmentation mask, by utilizing SwinTransformer as the encoder, employing a series of convolution-based blocks as the decoder. It is designed as a flexible framework, in which the upsample scale is adjustable and different backbones can be plugged in as the encoder. We set the downsampled 
LGE image ($64\times64$) as the training input, and the original segmentation mask ($512\times512$) as output. To keep more detailed information, the patch size is set as 2.

\begin{figure}[htb]
\vspace{-4ex}
\centerline{\includegraphics[width=0.48\textwidth]{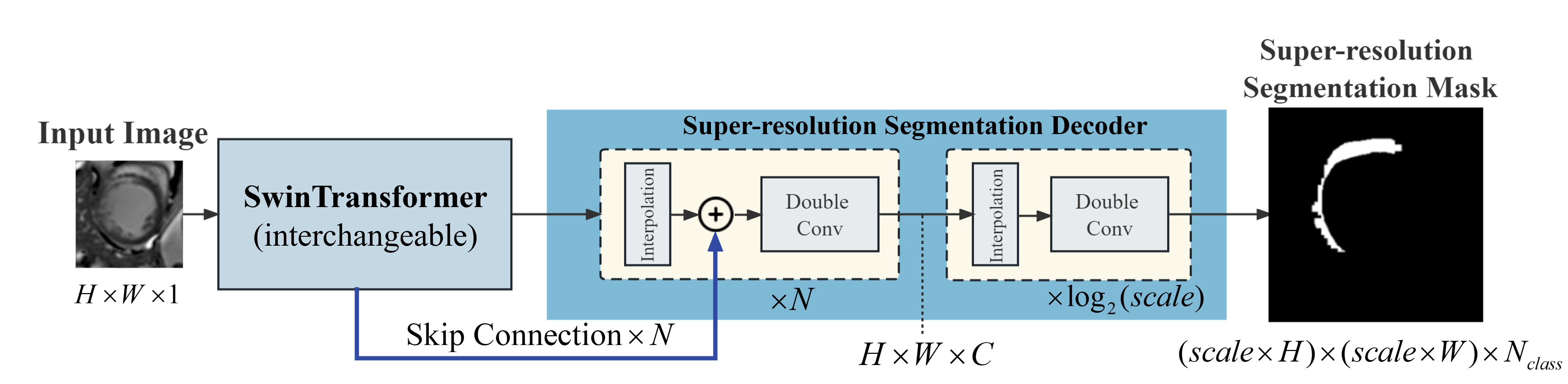}}
\vspace{-1.5ex}
\caption{The architecture of the SwinSSegNet. $scale$ stands for the upsampling scale. $H \times W$ and $C$ represents the origianl spatial shape and the channel number of the feature map respectively. The decoder of the SwinSSegNet can be divided into two parts. The first part focuses on upsampling the feature map to match the spatial dimensions of the input image. The subsequent part is responsible for generating the upsampled output segmentation mask. }
\label{fig:SwinSSegNet}
\vspace{-1ex}
\end{figure}

As shown in Fig. \ref{fig:SwinSSegNet}, within each layer of the first part, skip connection enables the transmission of detailed information from the encoder to the decoder. The second part has the flexibility to adjust the upsample scale. Each decoder block comprises an interpolation upsampling layer followed by a double convolution layer. The double convolution layer consists of two stacks, encompassing convolution, Rectified Linear Unit, and Batch Normalization layers.

\subsection{Universal Multi-task Self-supervised Learning Pre-training}

\begin{figure}[h]
\vspace{-1ex}
\centerline{\includegraphics[width=0.41\textwidth]{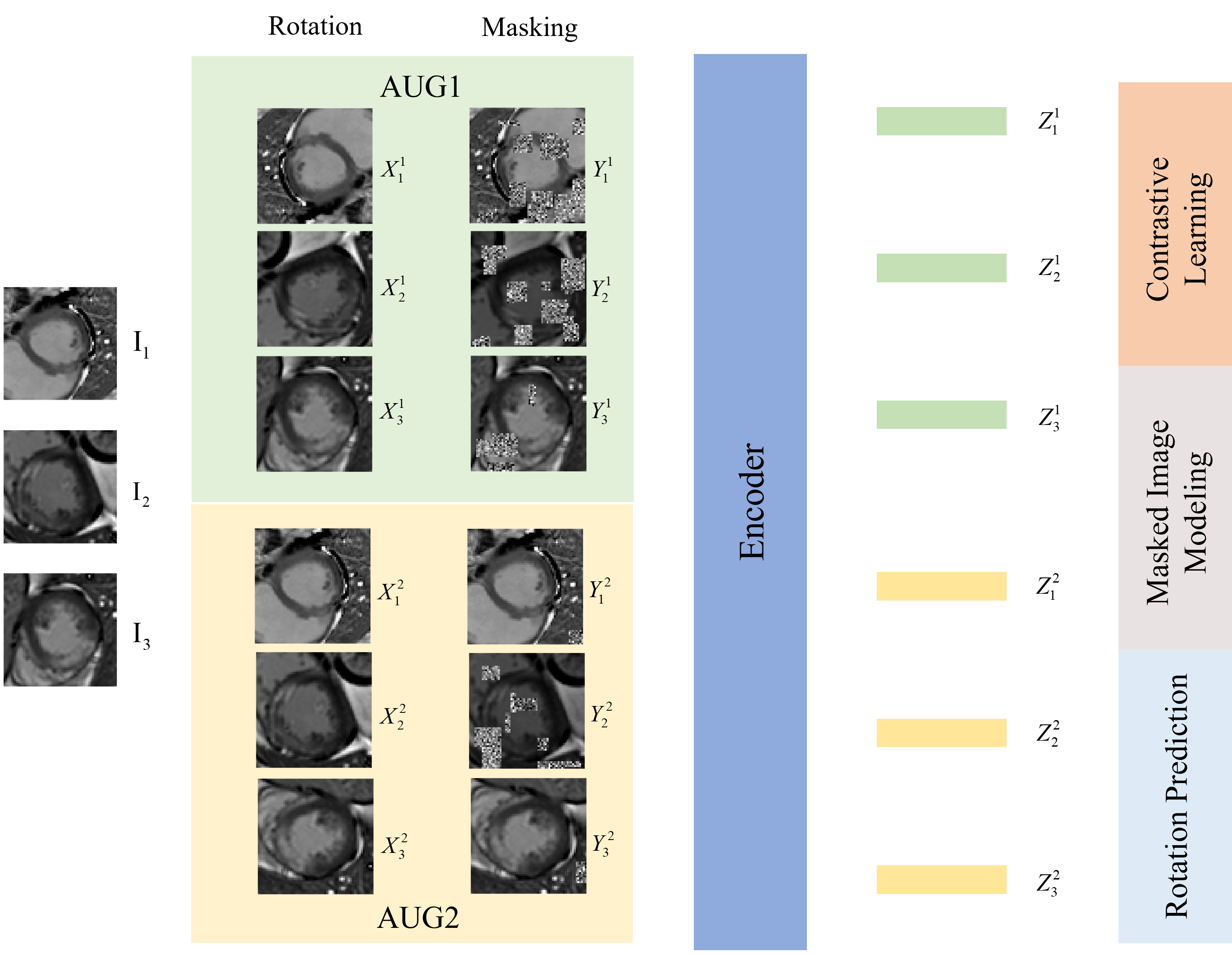}}
\vspace{-1ex}
\caption{The 2D mutli-task self-supervised learning framework. Input late gadolinium enhancement images are augmented with rotation and random masking, subsequently fed to the encoder as input.}
\label{fig:SSL}

\vspace{-2ex}
\end{figure}

Motivated by the recent work on 3D patch-wise SSL pre-training framework for Swin UNETR \cite{Tang2021SelfSupervisedPO}, we adopt a 2D SSL pre-training strategy to fully harness the capabilities of the SwinTransformer encoder. This strategy encompasses various components, including contrastive learning, which aims to encourage the model to capture general semantic features; masked image modeling, which assists the model in learning detailed features; and rotation prediction, which promotes the acquisition of spatial features. 
Additionally, we collect and crop LGE images from the ACDC and LiVScar datasets, incorporating them alongside our private dataset to form a hybrid dataset. Subsequently, the SSL pre-training is performed on this hybrid dataset, leading to the formulation of a novel strategy termed universal multi-task self-supervised learning  (U-SSL) pretraining strategy.

In the implementation,  the same data augmentation method is utilized to generate similar/dissimilar pairs for contrastive learning, masked images for masked image modeling, and rotated images for the rotation prediction task. As shown in Fig. \ref{fig:SSL}, the SSL data engine contains two sets of augmentation operations, AUG1 and AUG2. Each set encompasses rotation and masking. To exemplify, within AUG1, a batch of images, ($I_1, I_2, ..., I_n$), initially undergo rotation to yield ($X^1_1, X^1_2, ..., X^1_n$), and then are randomly masked out to produce ($Y^1_1, Y^1_2, ..., Y^1_n$). The rotation prediction task predicts $\hat{y_r}$, the rotation angle of ($I_i, X_i$), and the associated loss, $Loss_{ROT}$, is designed as \eqref{eq:loss_rot}, where ground truth $y_r$ $\in$ {$0^\circ$, $90^\circ$, $180^\circ$, $270^\circ$}. The masked image modeling task aims at generating $I_i$ from $Z_i$, and the loss, $Loss_{MIM}$, is designed as \eqref{eq:loss_MIR}. Contrastive learning task maximizes the dot product similarity (sim) between positive embedding pairs ($Z^1_i, Z^2_i$), while minimizing that between the other negative embedding pairs, and the loss, $Loss_{CL}$, is designed as \eqref{eq:loss_CL}.

\vspace{-2ex}
\begin{equation}
    Loss_{ROT}=\sum_{r=0}^{3} y_rlog({\hat{y}_r }),
    \label{eq:loss_rot}
\end{equation}

\vspace{-2ex}
\begin{equation}
    Loss_{MIM} =\hspace{0.3ex} \parallel X_i-I_i \parallel _1,  
    \label{eq:loss_MIR}
\end{equation}

\vspace{-3ex}
\begin{equation}
     Loss_{CL} = -log\frac{exp(\sum_{i}^{2N} sim(Z_i^1,Z_i^2)/t)}{ {\textstyle \sum_{i}^{2N}\sum_{k}^{2N}}1_{k\ne i}exp(sim(Z_i^1, Z_k^2)/t)},
    \label{eq:loss_CL}
\end{equation}
where t is the measurement of normalized temperature scale. $1_{k \ne i}$ is the indicator function evaluating to 1 if $k \ne i$.
\vspace{1ex}
\section{Experiments}
\subsection{Dataset}

Our private dataset consists of 277 unlabeled DTI images and 271 labeled LGE images, with the resolution of $64\times64$ and $512\times512$ respectively. Among these, 77 DTI-LGE image pairs share identical segmentation masks. We denote the 271 labeled LGE images as the LGE sub-dataset, and the 77 DTI-LGE image pairs with same segmentation masks as paired DTI-LGE image sub-dataset. Both the LGE image and DTI image have the target subject, MI. Manual segmentation of MI was undertaken by a CMR physicist with over 3 years of experience; these segmentations serve as the ground truth for the training and assessment of our proposed SwinSSegNet and pipeline. Additionally, We incorporate images from the ACDC and LiVScar datasets into our LGE subset to construct a hybrid LGE dataset for U-SSL pre-training. The ACDC dataset comprises 100 cine MRI scans (1,902 slices), and the LiVScar contains 30 images (200 slices). From each slice, the region of interest is cropped and upsampled to the resolution of $512\times512$ for our U-SSL pre-training process. 

\subsection{Implementation and Evaluation Details}

We conducted our experiments on an NVIDIA RTX3090 GPU with 24GB GPU RAM. The masking rate was set as 45\% for the SSL pretraining strategy. The AdamW optimizer was used. The CycleGANSeg was trained on our paired DTI-LGE image sub-dataset. The SwinSSegNet was pre-trained on the hybrid LGE dataset and fine-tuned on our LGE sub-dataset. All training procedures adopted the batch size of 24. 

With respect to the SwinSSegNet, the  Dice similarity coefficient \cite{dice1945measures} (DSC), indicative of the congruity between the segmentation outcome and the ground truth segmentation mask of the LGE image, was employed as the evaluation metric. In the context of the pipeline, the DSC between the segmentation result of the LGE style DTI image and the ground truth segmentation mask was utilized as the evaluation metric.

\subsection{Results of the SwinSSegNet}

\begin{figure}[htbp]
\vspace{-1ex}
\centerline{\includegraphics[width=0.5\textwidth]{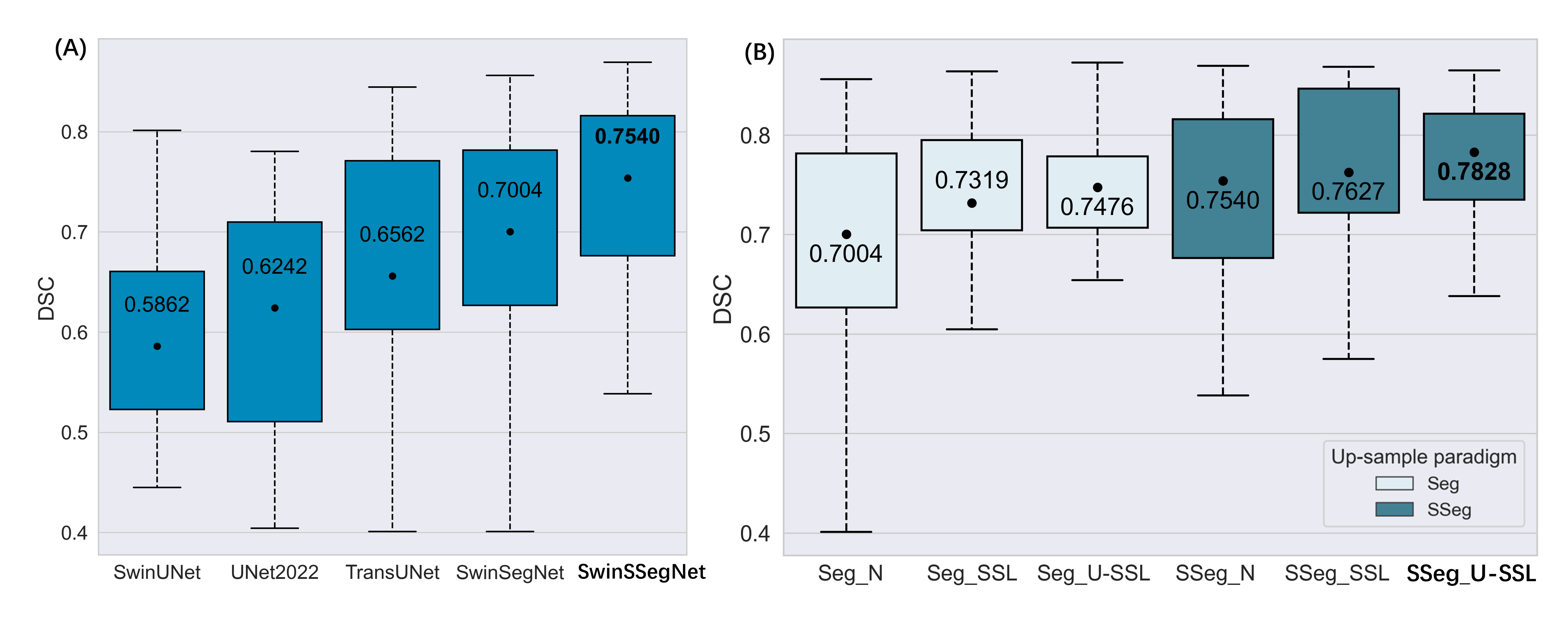}}
\vspace{-2ex}
\caption{(A). Comparison of the SwinSSegNet with baselines. SwinSegNet indicates adopting the upsample scale as 1 for the SwinSSegNet, and using interpolation to upsample the segmentation mask. (B). Comparison of the up-sample methods and the different pre-training strategies. SSeg stands for the end-to-end paradigm and Seg stands for the 2-stage paradigm (segmentation followed by bilinear interpolation upsampling). ``\_N", ``\_SSL", and ``\_U-SSL" represent trained from scratch, trained with multi-task self-supervised pre-training strategy, and trained with universal multi-task self-supervised pre-training strategy, respectively.  }
\label{fig:Compare_with_other_models}
\vspace{-1.8ex}
\end{figure}

We compared our proposed SwinSSegNet with state-of-the-art segmentation baselines (SwinUNet \cite{Cao2021SwinUnetUP}, UNet-2022 \cite{guo2022unet}, and TransUNet \cite{Chen2021TransUNetTM}) on the LGE sub-dataset (Fig. \ref{fig:Compare_with_other_models}.(A)). To ensure fairness, all models were trained from scratch. Compared with other models, the proposed SwinSSegNet achieves the best performance. 

Furthermore, a series of ablation studies were conducted to investigate various aspects of the proposed methodology. The performance of different up-sample paradigms and the performance of different pre-training strategies are compared in Fig. \ref{fig:Compare_with_other_models}.(B). The result indicates that the SSeg models are superior to the Seg models. Moreover, the employment of the SSL leads to enhanced model performance, and the incorporation of the public data can further boost the enhancement. A noteworthy observation is the decreasing standard deviation across model trained from scratch, model pre-trained with SSL, and model pre-trained with U-SSL. This observation highlights the substantial impact of contrastive learning, which facilitates the acquisition of comprehensive representation knowledge. Consequently, the model becomes adept at effectively handling segmentation tasks, even for challenging samples that initially exhibited subpar performance. Moreover, we compared the effect of choosing different backbone models as the encoder, the result is shown in Table. \ref{tab:encoder}. 

\begin{table}[htb]
\centering
\resizebox{0.45\textwidth}{!}{
\begin{tabular}{|c|cccl|}
\hline
Encoder           & ResNet \cite{He2015DeepRL}             & ConvNext \cite{Liu2022ACF}           & Biformer \cite{Zhu2023BiFormerVT}          & SwinTransformer    \\ \hline
DSC               & 0.718$\pm$0.0094     & 0.732$\pm$0.0093     & 0.745$\pm$0.0093     & 0.755$\pm$0.0090     \\ \hline
\end{tabular}}

\vspace{1ex}
\caption{Comparison of using different backbone model as encoder.}
\label{tab:encoder}
\vspace{-4.2ex}
\end{table}

In the encoder comparison, we adopted the patch size as 2 for Biformer and SwinTransformer, and the downsample size as 2 for the first pooling layer in ResNet and ConvNext. The result indicates that SwinTransformer stands out as the most effective encoder.

\subsection{Results of the Pipeline}

To illustrate the superiority of our proposed pipeline (CycleGANSeg+SSeg), i.e., introducing the CycleGANSeg to convert DTI image to LGE style before using the SwinSSegNet trained with U-SSL pre-training strategy to perform segmentation, we compared the cases of not using style transfer model (DTI SSeg) and using the original CycleGAN as the style transfer model (CycleGAN+SSeg). 

\begin{figure}[htbp]
\vspace{-0ex}
\centerline{\includegraphics[width=0.32\textwidth]{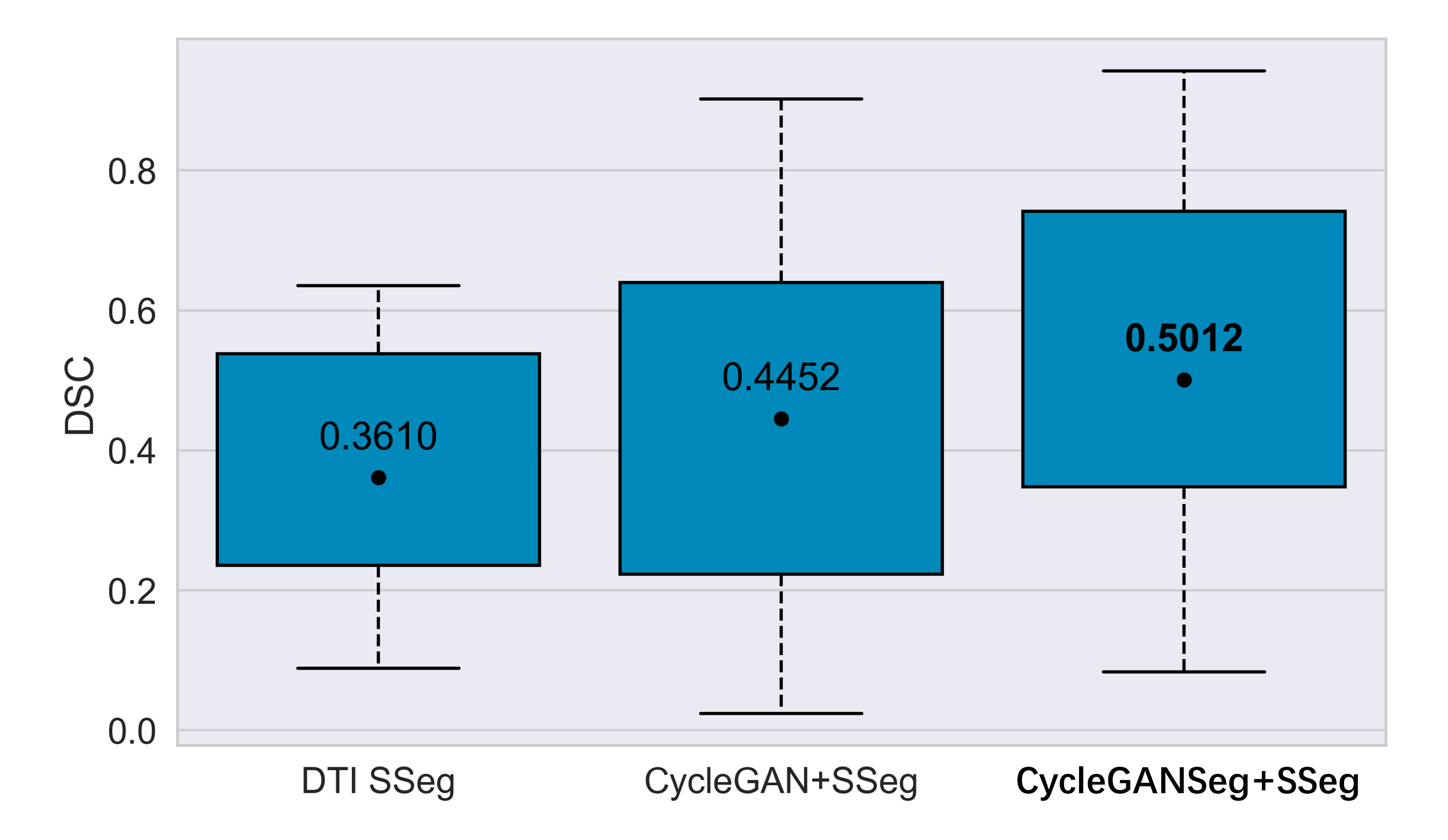}}
\vspace{-2ex}
\caption{Comparison of different pipelines.  DTI SSeg stands for the SwinSSegNet trained from scratch on DTI image data. CycleGAN+SSeg stands for using the original CycleGAN as the style transfer model, and the SwinSSegNet trained with U-SSL pre-training strategy on LGE data as the SSeg model.}
\label{fig:style}
\vspace{-1ex}
\end{figure}

\begin{figure}[htbp]
\vspace{-2ex}
\centerline{\includegraphics[width=0.48\textwidth]{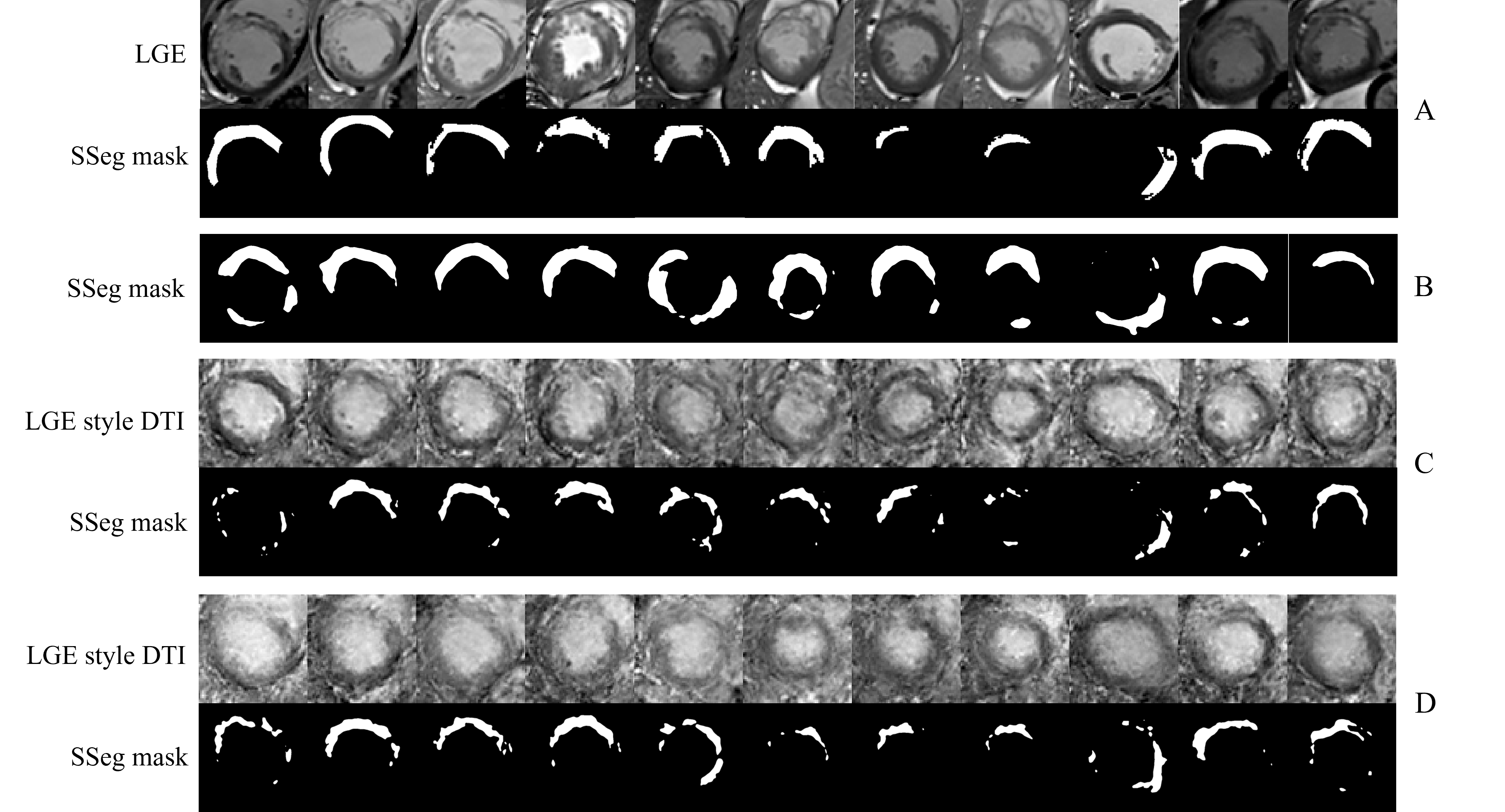}}
\vspace{-1.2ex}
\caption{Visualized results of the different pipelines. Row A illustrates the ground truth LGE image and SSeg mask. Row B illustrates the SSeg results of DTI SSeg model. Row C and Row D illustrate the LGE style DTI image, SSeg results of CycleGAN+SSeg pipeline and CycleGANSeg+SSeg pipeline respectively.}
\label{fig:style_results}
\vspace{-2ex}

\end{figure}

The quantitative result shown in Fig. \ref{fig:style} and the visualized results Fig. \ref{fig:style_results} indicate that our pipeline effectively highlights the MI area and achieves the best MI segmentation performance. The DTI SSeg model demonstrates poor performance, while CycleGAN+SSeg improves performance but introduces increased variance. This aligns with expectations, as the original CycleGAN lacks a dedicated mechanism to preserve the underlying segmentation mask. In contrast, the proposed CycleGANSeg+SSeg enhances the average performance while mitigating the range of both upper and lower bounds. This is achieved through the integration of the Segmentor in training process, which effectively helps the model to preserve the original segmentation mask, thereby stabilizing the style transfer process.

\section{Conclusions}
We present a novel pipeline for MI super-resolution segmentation on DTI image, incorporating the CycleGANSeg and the SwinSSegNet. The CycleGANSeg transforms DTI image to LGE style, bridges domain gaps, and simplifies MI segmentation. The SwinSSegNet surpasses the two-step segmentation paradigm in generating high-resolution segmentation mask, and the integration of the U-SSL pre-training strategy further enhances the segmentation performance.

\newpage

\bibliographystyle{IEEEtran}
\bibliography{b}

\end{document}